\documentclass[aps,pre,reprint,superscriptaddress,nofootinbib,floatfix]{revtex4-2}

\usepackage{amsmath,amssymb,amsfonts}
\usepackage{bm}
\usepackage{graphicx}
\usepackage{booktabs}
\usepackage{hyperref}
\hypersetup{colorlinks=true,linkcolor=black,citecolor=black,urlcolor=black}

\begin{document}

\title{Pole--Zero Geometry, Model Reduction, and Identifiability in Sensory Adaptation}

\author{Gunn Kim}
\email{gunnkim@sejong.ac.kr}
\affiliation{Department of Physics, Sejong University, Seoul 05006, Republic of Korea}

\date{\today}

\begin{abstract}
Sensory adaptation provides a concrete setting in which low-order system identification can fail qualitatively. We show that one fixed higher-order adaptive system composed entirely of real first-order relaxation modes can be reduced to opposite sides of the second-order pole boundary: low-frequency moment matching gives $\rho_{\rm moment}=4.50$, whereas finite-window fitting gives $\rho_{\rm window}=3.31$, and the inferred pole class changes further with sampling protocol. Thus the real-versus-complex classification of a reduced model is not itself reduction invariant. We then use the general two-state spectrum to connect stochastic identifiability to adaptation: for nontrivial coupling and one-state observation, cross diffusion drops out of the scalar spectrum when the hidden state has no self-relaxation. In the adaptive model, this condition is precisely the integral-memory limit that produces exact adaptation, while leaky memory restores spectral sensitivity. For the exact-adaptation model, the Gaussian path-space irreversibility nevertheless depends on the hidden cross-diffusion channel. Hence $\{H,S_x\}$ does not determine the irreversibility rate. Independently, for a specified all-even reduced two-state drift with $\rho<4$, the drift-only lower bound is $\sigma \ge \tau_x^{-1}(4/\rho-1)$. Published \textit{E.~coli} and \textit{C.~elegans} responses provide biological examples of these limits. The distinction established here between transfer-function invariants, reduction-dependent properties, and hidden-state quantities provides a concrete framework for evaluating the limitations of low-dimensional models of adaptive biological dynamics.
\end{abstract}

\maketitle

\section{Introduction}

Sensory adaptation allows biological systems to respond to changes in a persistent environment rather than to the persistent signal itself. In bacterial chemotaxis, classic tethered-cell experiments revealed biphasic temporal responses in which recent and earlier stimulus histories enter with opposite signs \cite{Macnab1972,Berg1975,Block1982,Segall1986}. At the biochemical level, robust return toward the prestimulus state motivated integral-feedback descriptions of the chemotaxis pathway \cite{Barkai1997,Alon1999,Yi2000,MelloTu2003,Tu2008}. More generally, adaptation can be implemented by distinct network architectures, including negative feedback and incoherent feedforward motifs \cite{Ma2009}. A similar diversity appears in \textit{Caenorhabditis elegans} (\textit{C. elegans}): the amphid wing C (AWC) olfactory sensory neurons exhibit established cell-intrinsic adaptation pathways \cite{Colbert1995,LEtoile2000,LEtoile2002}, and their olfactory responses show biphasic temporal filtering \cite{Chalasani2007,Chalasani2010,Kato2014}. Antagonistic parallel pathways and adaptive-threshold dynamics provide distinct descriptions of those temporal responses \cite{Kato2014,Levy2020}.

These studies establish how adaptation can be generated, but leave a different question largely open: \emph{what aspects of the inferred dynamics are actually determined by the measurements?} Different internal realizations can produce the same input--output response, while fitting a higher-order system with a low-order model can introduce dynamical features absent from the underlying relaxation spectrum. A recent preprint by Browning \textit{et al.} analyzes structural identifiability under partial observation for stochastic differential equations, including linear two-state Ornstein--Uhlenbeck processes \cite{Browning2025}. Partial observation creates a related thermodynamic problem, because hidden correlations can affect nonequilibrium irreversibility without appearing in a measured scalar fluctuation spectrum \cite{Martinez2019}. Thus a successful low-dimensional fit need not uniquely identify either the microscopic mechanism or every qualitative property attributed to the fitted dynamics.

In this study, we address these ambiguities using transfer-function and pole--zero analysis, explicit higher-order realizations, systematic model reduction, and stochastic Ornstein--Uhlenbeck theory \cite{Uhlenbeck1930}. We use antagonistic parallel subtraction as a realization example, not as a newly proposed biological circuit. Our central deterministic result is that one fixed higher-order system containing only real relaxation modes gives $\rho_{\rm moment}=4.50$ under low-frequency moment matching but $\rho_{\rm window}=3.31$ under finite-window fitting, reversing the real-versus-complex pole classification of the reduced model. For a general real two-state OU process, the scalar spectrum makes the condition transparent: for nontrivial coupling, cross diffusion drops out precisely when the hidden state has no self-relaxation. This structure is consistent with recent structural-identifiability results for partially observed OU systems \cite{Browning2025}. In the adaptive model it acquires a direct physical interpretation, because the same condition is the integral-memory limit that produces exact adaptation. Published \textit{E.~coli} and \textit{C.~elegans} responses provide biological examples of these limits. The results distinguish transfer-function invariants from properties introduced by model reduction or lost through partial observation, clarifying what a successful low-order model can and cannot reveal about adaptive biological dynamics.

\section{Realization-invariant adaptive transfer function}

We begin with normalized deviations of a sensory output $x(t)$, a memory variable $m(t)$, and a stimulus $s(t)$,
\begin{align}
 \tau_x \dot x &= -x-m+g s(t), \label{eq:x}\\
 \tau_m \dot m &= x-\kappa m, \label{eq:m}
\end{align}
where $\tau_x,\tau_m>0$, $g$ is a small-signal gain, and $\kappa\ge0$ is a dimensionless leakage parameter. The limit $\kappa=0$ is the ideal-integrator realization. For a constant input $s_0$,
\begin{equation}
 x_{\rm ss}=g s_0\frac{\kappa}{1+\kappa},\qquad
 m_{\rm ss}=g s_0\frac{1}{1+\kappa}. \label{eq:ss}
\end{equation}
Thus exact return of the observed output occurs only at $\kappa=0$.

For Laplace variable $p$,
\begin{equation}
 H_2(p)\equiv\frac{X(p)}{S(p)}=
 g\frac{\tau_m p+\kappa}
 {\tau_x\tau_m p^2+(\tau_m+\kappa\tau_x)p+(1+\kappa)}.
 \label{eq:Hgeneral}
\end{equation}
The zero and dc gain are
\begin{equation}
 z=-\frac{\kappa}{\tau_m},\qquad
 H_2(0)=g\frac{\kappa}{1+\kappa}. \label{eq:zeroH0}
\end{equation}
This is the first separation used throughout the paper: zero placement determines the completeness of dc rejection, whereas the poles determine transient relaxation. In statistical-physics language, the poles are the decay eigenvalues of the linearized dynamics, while a zero marks a cancellation in the measured susceptibility \cite{Kubo1966}: internal variables can continue to evolve even when a particular stimulus component leaves no signal in the observed output. The pole--zero description therefore separates relaxation in state space from information removed by the observation map.

Defining
\begin{equation}
 \rho=\frac{\tau_m}{\tau_x}, \label{eq:rho}
\end{equation}
the poles coalesce when
\begin{equation}
 \rho_{\rm EP}^{\pm}=\left(\sqrt{1+\kappa}\pm1\right)^2. \label{eq:EP}
\end{equation}
On the exact-integrator slice $\kappa=0$,
\begin{equation}
 H_2(p)=g\frac{\tau_m p}{\tau_x\tau_m p^2+\tau_m p+1}, \label{eq:Hideal}
\end{equation}
and
\begin{equation}
 \begin{cases}
 \rho<4 &: \text{complex poles},\\
 \rho=4 &: \text{double pole},\\
 \rho>4 &: \text{two real poles}.
 \end{cases} \label{eq:regimes}
\end{equation}
The damping ratio is $\zeta=\sqrt{\rho}/2$. The value four is therefore not a universal biological constant: it is the upper exceptional-point intersection of the $\kappa=0$ slice.

\section{Parallel subtraction as a realization of adaptation}

The exact-integrator model has a particularly transparent realization when $\rho>4$. Its denominator can be factored as
\begin{equation}
 \tau_x\tau_m p^2+\tau_m p+1=(1+pT_f)(1+pT_s), \label{eq:factor}
\end{equation}
where
\begin{align}
 T_{f,s}&=\frac{\tau_m}{2}\left(1\mp\sqrt{1-\frac{4}{\rho}}\right),\nonumber\\
 T_fT_s&=\tau_x\tau_m,\qquad T_f+T_s=\tau_m. \label{eq:Tfs}
\end{align}
Equation~(\ref{eq:Hideal}) then becomes
\begin{equation}
 H_2(p)=\frac{g\tau_m}{T_s-T_f}
 \left[\frac{1}{1+pT_f}-\frac{1}{1+pT_s}\right]. \label{eq:parallel2}
\end{equation}
Thus the same transfer function can be viewed either as integral feedback or as the subtraction of two low-pass temporal branches. The origin zero is simply equal dc cancellation: both branches transmit a constant input with the same steady gain, and their difference vanishes. This representation is directly reminiscent of the positive and negative temporal weights measured by Segall, Block, and Berg \cite{Segall1986} and of incoherent feedforward adaptation \cite{Ma2009}.

The representation also sharpens the interpretation of $\rho=4$. For $\rho>4$, two distinct real branch times exist. At $\rho=4$ they coalesce, giving the repeated-pole limit
\begin{equation}
 H_2(p)=\frac{2g t_0 p}{(1+p t_0)^2},\qquad
 t_0=\sqrt{\tau_x\tau_m}. \label{eq:EPlimit}
\end{equation}
For $\rho<4$, no decomposition into two stable real first-order branches exists. A fitted complex pair is therefore a statement about the chosen second-order reduction, not proof that the microscopic biology contains an oscillatory two-state module.

A natural extension is to add a common upstream stage before the two antagonistic branches,
\begin{equation}
 H_{\parallel}(p)=K G_a(p)\left[G_f(p)-\alpha G_s(p)\right], \label{eq:Hparallel}
\end{equation}
with
\begin{equation}
 G_j(p)=\frac{1}{1+pT_j}. \label{eq:Gj}
\end{equation}
The parameter $\alpha$ is the relative strength of the slow subtractive path. Exact adaptation corresponds to $\alpha=1$. For unequal branch gains,
\begin{equation}
 H_{\parallel}(0)=K(1-\alpha),\qquad
 z_{\parallel}=-\frac{1-\alpha}{T_s-\alpha T_f}. \label{eq:branchzero}
\end{equation}
Thus a finite response zero can be read either as leaky memory in Eqs.~(\ref{eq:x})--(\ref{eq:m}) or, in a feedforward realization, as imperfect cancellation of two pathways.

Kato \textit{et al.} used precisely the structural form ``common stage $\rightarrow$ fast and slow opposite-sign branches'' to describe AWC temporal filters \cite{Kato2014}. Motivated by the data fits below, we consider the repeated-fast-pole specialization $T_a=T_f\equiv T$,
\begin{equation}
 H_3(p)=K\frac{p(T_s-T)}{(1+pT)^2(1+pT_s)} \qquad (\alpha=1). \label{eq:H3}
\end{equation}
It has three real poles, an origin zero, and relative degree two. The relative degree---the number of poles in excess of zeros---has a direct physical meaning: it fixes the high-frequency roll-off and therefore how abruptly the measured output can respond at short times. Consequently its impulse response starts from zero, $h(0^+)=0$, whereas the minimal two-state model has relative degree one and $h(0^+)=g/\tau_x$. In this sense an extra unit of relative degree acts as an additional unresolved lag that smooths rapid input changes. The additional common stage therefore changes short-time peak localization without changing the exact-adaptation condition.

\section{Time- and frequency-domain consequences}

\begin{figure}[t]
\centering
\includegraphics[width=\linewidth]{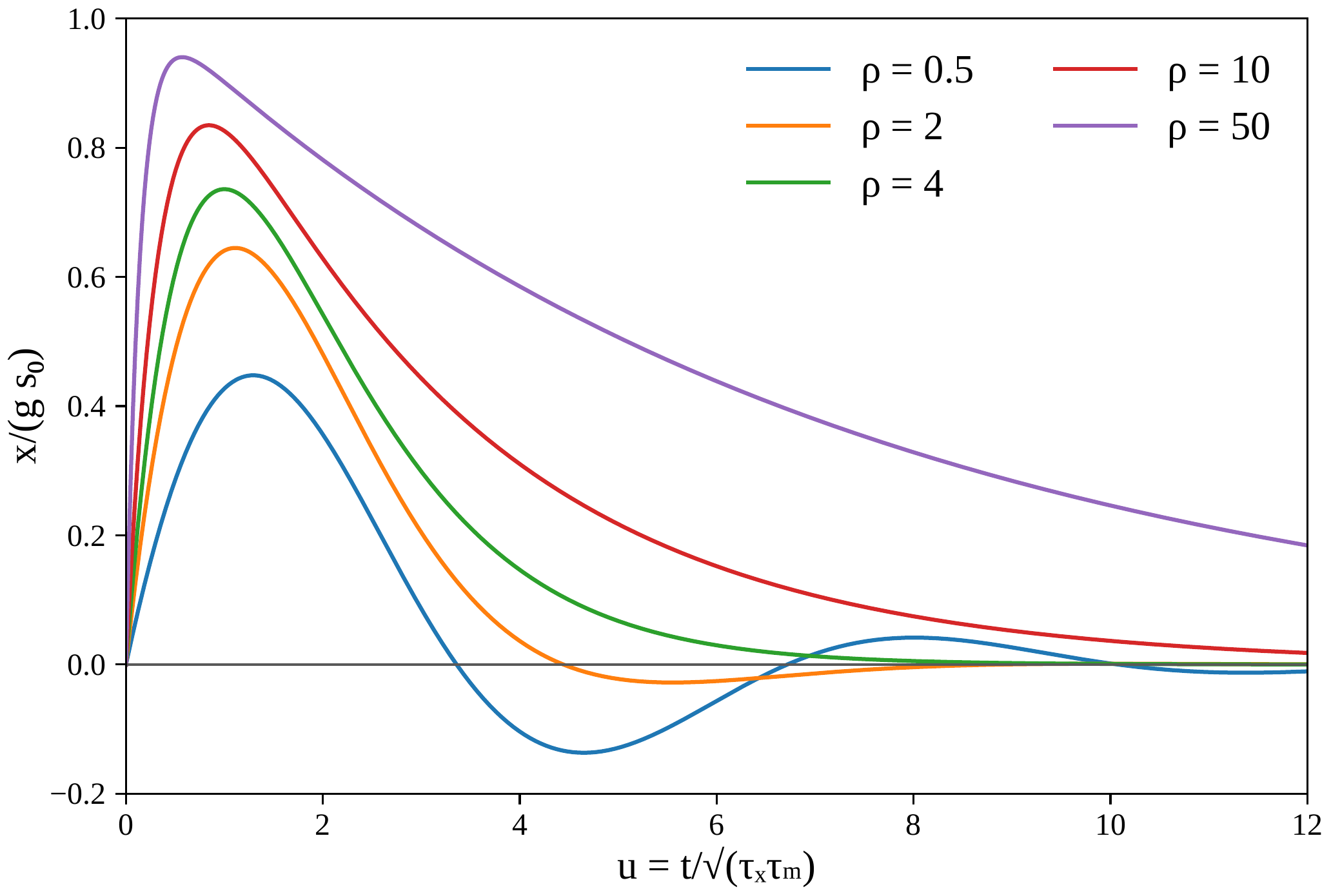}
\caption{Normalized response to a persistent step on the exact-integrator slice $\kappa=0$ for several values of $\rho=\tau_m/\tau_x$. Time is measured in units $t_0=\sqrt{\tau_x\tau_m}$ and the output in units of $gs_0$. In the parallel-subtraction interpretation, a persistent input excites two temporal branches whose transient mismatch produces the pulse while equal dc gains cancel at long times. The poles control the transient morphology: complex reduced modes occur for $\rho<4$, the poles coalesce at $\rho=4$, and two real modes occur for $\rho>4$.}
\label{fig:step}
\end{figure}

Figure~\ref{fig:step} shows the normalized step response of the minimal exact-integrator slice. In the parallel picture, the pulse is the temporary mismatch between fast and slow branches after both receive the same persistent input. They are equal at long times, so the output returns to baseline. The pole locations determine whether that mismatch relaxes through an effective complex pair or through distinct real decays.

\begin{figure}[t]
\centering
\includegraphics[width=\linewidth]{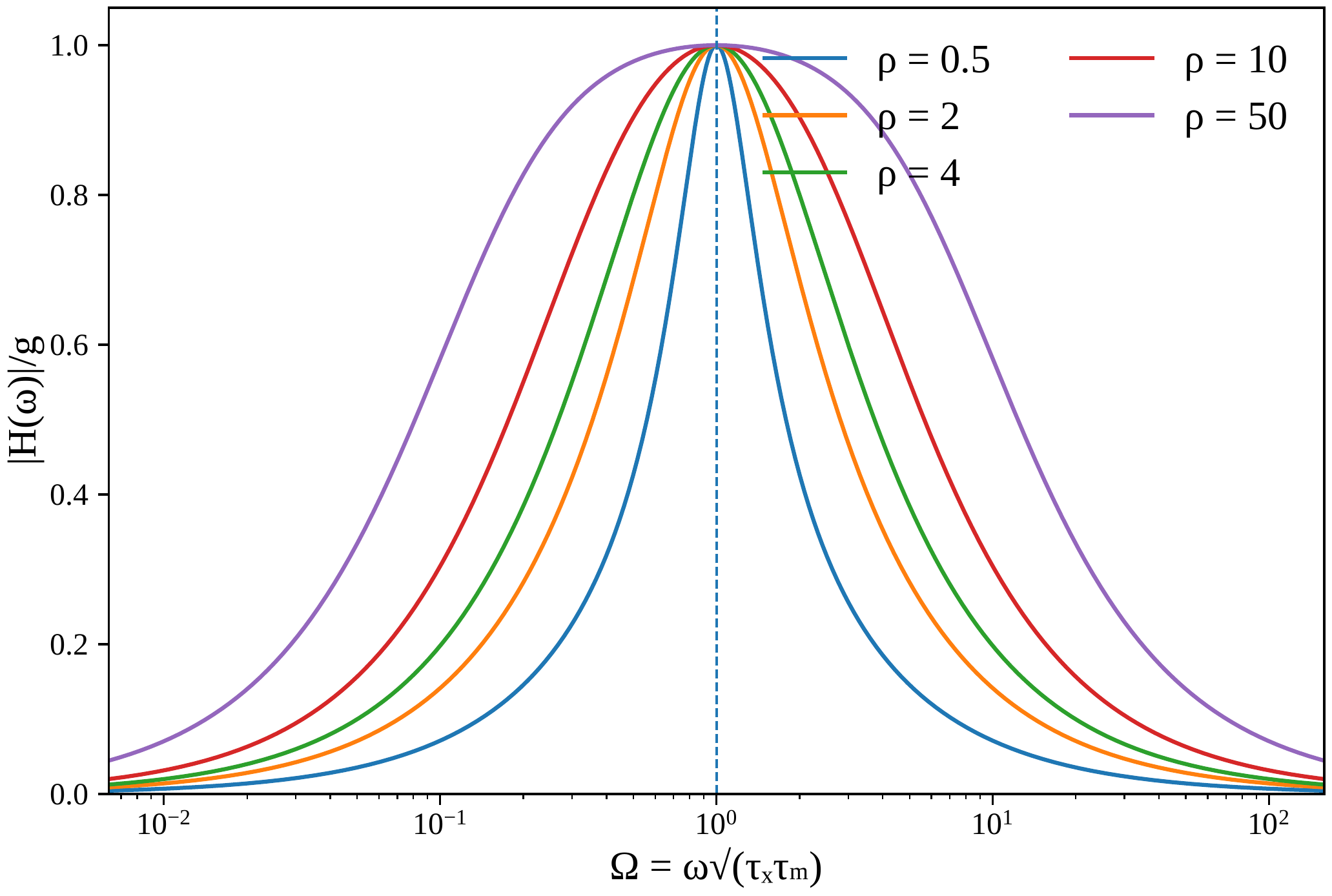}
\caption{Frequency response on the exact-integrator slice. With $\Omega=\omega\sqrt{\tau_x\tau_m}$, all minimal-model curves peak at $\Omega=1$. Parallel subtraction suppresses the common dc component, while finite branch kinetics suppress sufficiently fast modulation; the surviving intermediate band is therefore the frequency-domain counterpart of temporal comparison. The minimal two-state model has $\omega_p=(\tau_x\tau_m)^{-1/2}$ and $\Delta\omega=1/\tau_x$. A common upstream stage, as in Eq.~(\ref{eq:H3}), adds a further high-frequency pole and breaks the exact log-frequency inversion symmetry.}
\label{fig:frequency}
\end{figure}

For $\kappa=0$, with $\Omega=\omega\sqrt{\tau_x\tau_m}$,
\begin{equation}
 \frac{|H_2(i\omega)|^2}{g^2}=
 \frac{\rho\Omega^2}{(1-\Omega^2)^2+\rho\Omega^2}. \label{eq:mag}
\end{equation}
The response obeys $|H(\Omega)|=|H(1/\Omega)|$, peaks at
\begin{equation}
 \omega_p=\frac{1}{\sqrt{\tau_x\tau_m}}, \label{eq:omegap}
\end{equation}
and has half-power bandwidth
\begin{equation}
 \Delta\omega=\frac{1}{\tau_x},\qquad Q=\frac{\omega_p}{\Delta\omega}=\rho^{-1/2}. \label{eq:bandwidth}
\end{equation}
Figure~\ref{fig:frequency} is therefore also a subtraction diagram in frequency space: sufficiently slow components are rejected because the two branches agree, sufficiently rapid components are suppressed by finite kinetics, and intermediate frequencies are transmitted because the branches have different phase and amplitude.

\section{Biological examples of realization and model order}

We compare the minimal two-state kernel and the repeated-fast-pole cascade--parallel kernel, Eq.~(\ref{eq:H3}), with the same graph-digitized response curves used previously. The fits use the reported stimulus onset times and are intended as coarse-grained system identification, not microscopic parameter estimation. Details and digitization sensitivity are given in the Supplemental Material (SM).

For tethered \textit{E.~coli}, the minimal fit gives
\begin{equation}
 \tau_x=0.529~{\rm s},\quad \tau_m=1.666~{\rm s},\quad
 \rho=3.148,\quad R^2=0.928. \label{eq:E2fit}
\end{equation}
The experimental peak is higher and narrower than this curve: the data reach a CCW bias of about $0.93$ at $1.8$ s, whereas the minimal fit peaks near $0.886$ at $1.98$ s. The balanced cascade--parallel fit yields
\begin{equation}
 T=0.268~{\rm s},\qquad T_s=1.364~{\rm s},\qquad
 R^2=0.958, \label{eq:E3fit}
\end{equation}
with a fitted peak near $0.905$ at $1.90$ s. Because this repeated-fast-pole form uses the same number of fitted parameters as the two-state fit (including the motor-bias baseline), the improvement is not obtained by adding a free timescale; $\Delta{\rm AIC}=-13.44$ and $\Delta{\rm BIC}=-13.44$ relative to the two-state fit. Allowing unequal branch gains is not favored by the information criteria, consistent with nearly balanced temporal subtraction in the bacterial response.

For the corrected AWC$^{\rm ON}$ calcium response, the minimal ideal-slice fit is
\begin{equation}
 \tau_x=3.139~{\rm s},\quad \tau_m=25.974~{\rm s},\quad
 \rho=8.275,\quad R^2=0.952. \label{eq:A2fit}
\end{equation}
It peaks at about $85.5\%$ near $17.9$ s, compared with the graph-digitized peak of about $95\%$ at $16$ s. The balanced cascade--parallel form instead gives
\begin{equation}
 T=1.514~{\rm s},\qquad T_s=24.56~{\rm s},\qquad
 R^2=0.989, \label{eq:A3fit}
\end{equation}
with a peak near $89.5\%$ at $16.85$ s. The parameter count is again unchanged, and the improvement is large: $\Delta{\rm AIC}=-30.32$ and $\Delta{\rm BIC}=-30.32$ relative to the two-state fit. Allowing branch imbalance gives $\alpha\simeq0.88$, $T\simeq1.64$ s, and $T_s\simeq15.9$ s with $R^2\simeq0.992$. The associated asymptotic offset is about $16\%$ and the zero is of order $z\simeq-8\times10^{-3}\,{\rm s}^{-1}$, corresponding to a zero timescale of roughly $120$ s. Because the digitized trace extends only to $60$ s, this finite-zero estimate should be regarded as suggestive rather than sharply identified.

\begin{figure*}[t]
\centering
\includegraphics[width=0.96\textwidth]{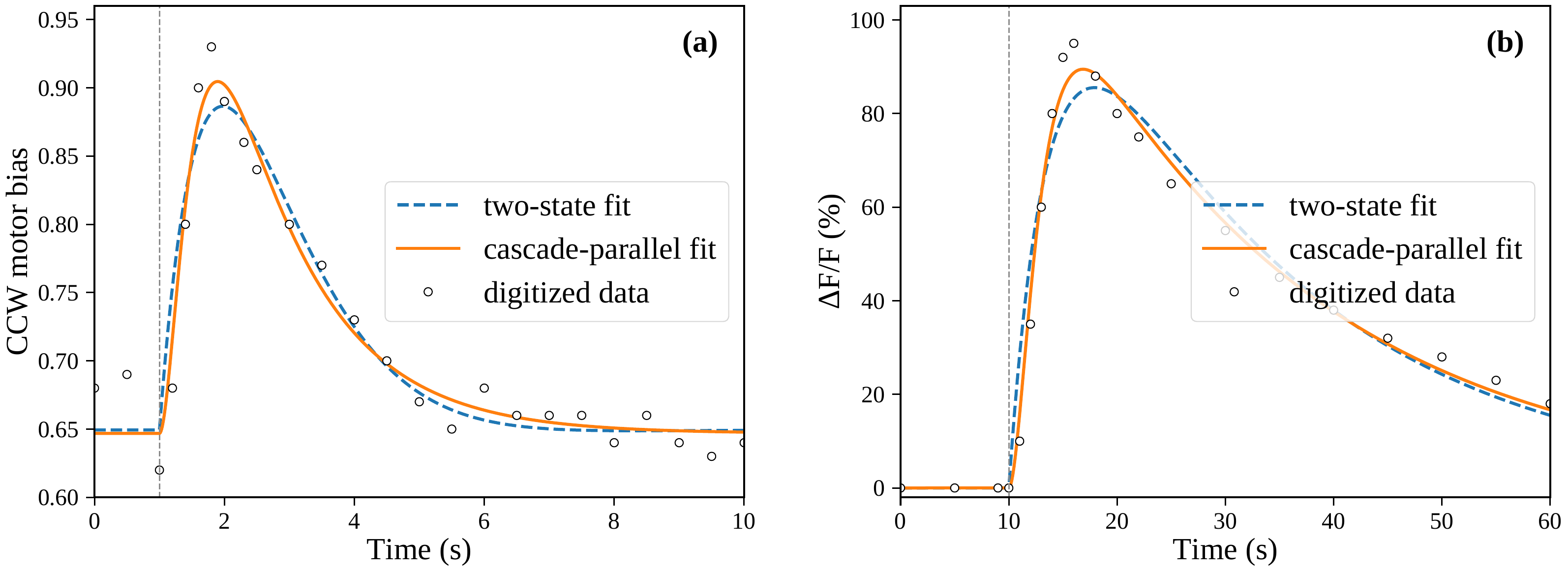}
\caption{Graph-digitized biological responses compared with two coarse-grained realizations. (a) Wild-type \textit{E.~coli} tethered-cell CCW motor bias from Segall, Block, and Berg~\cite{Segall1986}. The dashed two-state fit has $\tau_x=0.529$ s, $\tau_m=1.666$ s, $\rho=3.148$, and $R^2=0.928$; the solid balanced cascade--parallel fit has $T=0.268$ s, $T_s=1.364$ s, and $R^2=0.958$. (b) Corrected AWC$^{\rm ON}$ calcium response after isoamyl-alcohol removal~\cite{Chalasani2016}. The dashed two-state fit has $\tau_x=3.139$ s, $\tau_m=25.974$ s, $\rho=8.275$, and $R^2=0.952$; the solid balanced cascade--parallel fit has $T=1.514$ s, $T_s=24.56$ s, and $R^2=0.989$. In both panels the additional common stage sharpens the response peak. Dashed vertical lines mark the reported stimulus transitions. The fits describe input--output dynamics and are used as biological illustrations of model order, not as unique microscopic reconstructions.}
\label{fig:biofits}
\end{figure*}

\section{Pole--zero geometry versus internal realization}

\begin{figure*}[t]
\centering
\includegraphics[width=0.98\textwidth]{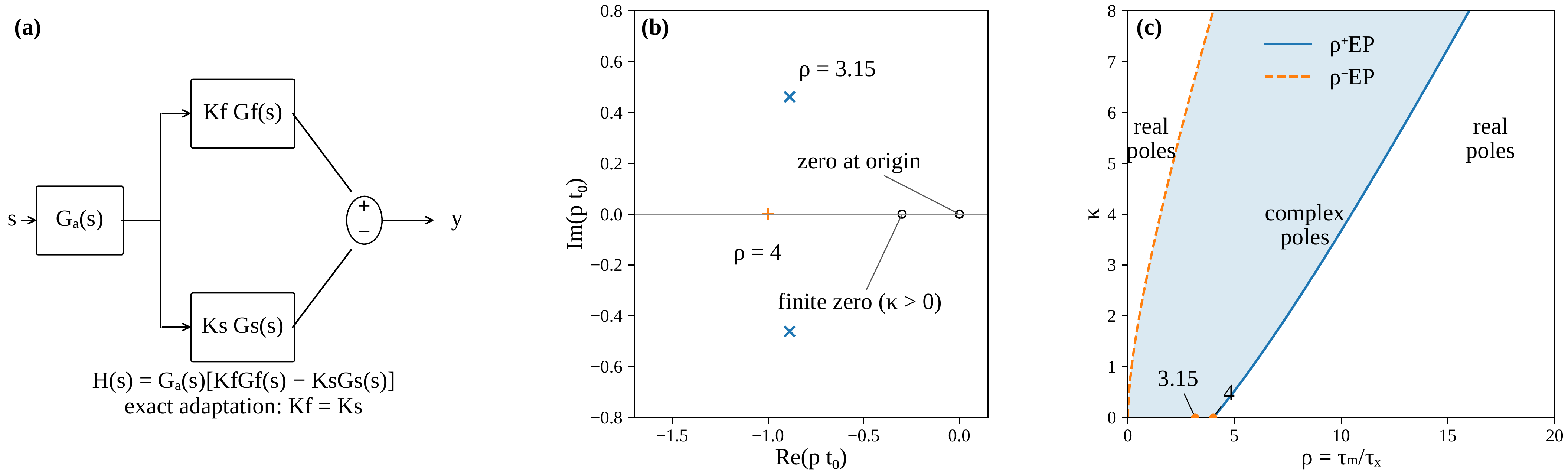}
\caption{Parallel realization and pole--zero classification. (a) A common front-end $G_a$ feeds fast and slow branches that are subtracted at the output. Exact adaptation corresponds to equal dc branch gains; unequal gains produce partial adaptation. (b) Minimal-model pole locations in units of $t_0^{-1}$ on the exact-integrator slice, shown for the reduced complex-pole example $\rho=3.15$, the double-pole boundary at $\rho=4$, and representative zeros at the origin and at finite $\kappa>0$. The complex pair at $\rho<4$ is a property of the reduced second-order representation and need not imply microscopic oscillatory states. (c) Pole regimes in the $(\rho,\kappa)$ plane. The exceptional-point branches $\rho_{\rm EP}^{\pm}=(\sqrt{1+\kappa}\pm1)^2$ bound the complex-pole region. The value $\rho=4$ is only the $\kappa=0$ upper-branch point, not a universal threshold.}
\label{fig:polezero}
\end{figure*}

Figure~\ref{fig:polezero} summarizes the distinction. Panel (a) displays the cascade--parallel architecture: a common stage feeds fast and slow branches that enter the output with opposite sign. Equal dc gains place a zero at the origin; imbalance moves the zero. Panel (b) shows the minimal-model poles and zero in the complex plane, and panel (c) shows the generalized exceptional-point boundaries. The pole--zero description is realization invariant, whereas labels such as ``feedback memory,'' ``fast branch,'' or ``slow branch'' refer to a particular internal representation.

This perspective also clarifies the role of relative degree. The minimal two-state transfer has one more pole than zero and therefore a nonzero instantaneous impulse limit. The cascade--parallel form has two more poles than zeros and therefore suppresses the immediate response. The sharper peaks in Fig.~\ref{fig:biofits} are consequently not evidence for stronger ``inertia''; they indicate that the measured channel contains at least one additional dynamical stage. In AWC, Kato \textit{et al.} already separated a neuronal filter from GCaMP dynamics and used an intrinsic three-variable model with a common stage followed by opposite-sign parallel paths \cite{Kato2014}. In \textit{E.~coli}, the corresponding extra stage may reside in receptor signaling, motor transduction, stimulus delivery, or a combination of these; the present motor-bias trace alone cannot assign it microscopically.

\section{Reduction-dependent pole classification}

The bacterial result has a further consequence. Model reduction here is a dynamical coarse graining: several relaxation modes are compressed into two effective modes, so the reduced parameters depend on which part of the response is required to survive that compression. Low-frequency moment matching preserves the first coefficients of the $p\to0$ denominator expansion and therefore emphasizes the slow relaxation structure. Finite-window fitting instead minimizes the error at the sampled times and therefore emphasizes the experimentally resolved transient, especially the peak and early relaxation. Comparing the two is thus a direct test of information loss under coarse graining rather than merely a comparison of fitting algorithms.

The best cascade--parallel fit contains only real poles, including a repeated fast pole. Two different reductions of this \emph{same} higher-order transfer function onto the same exact-adaptation two-state family nevertheless give opposite qualitative pole classes. In both reductions the zero is fixed at the origin; what differs is how the two denominator parameters are selected. Matching the low-frequency denominator through second order gives
\begin{equation}
 \rho_{\rm moment}=4.50>4, \label{eq:rhomoment}
\end{equation}
whereas sampling the noiseless response at the graph-digitized bacterial time points and minimizing the time-domain residual gives
\begin{equation}
 \rho_{\rm window}=3.31<4. \label{eq:projection}
\end{equation}
The two procedures retain different information---the $p\to0$ relaxation expansion versus the sampled time-domain transient---so different numerical parameters are not by themselves surprising. The nontrivial point is qualitative: information discarded by the reduction is sufficient to reverse the inferred real-versus-complex pole class.

\begin{figure*}[t]
\centering
\includegraphics[width=0.90\textwidth]{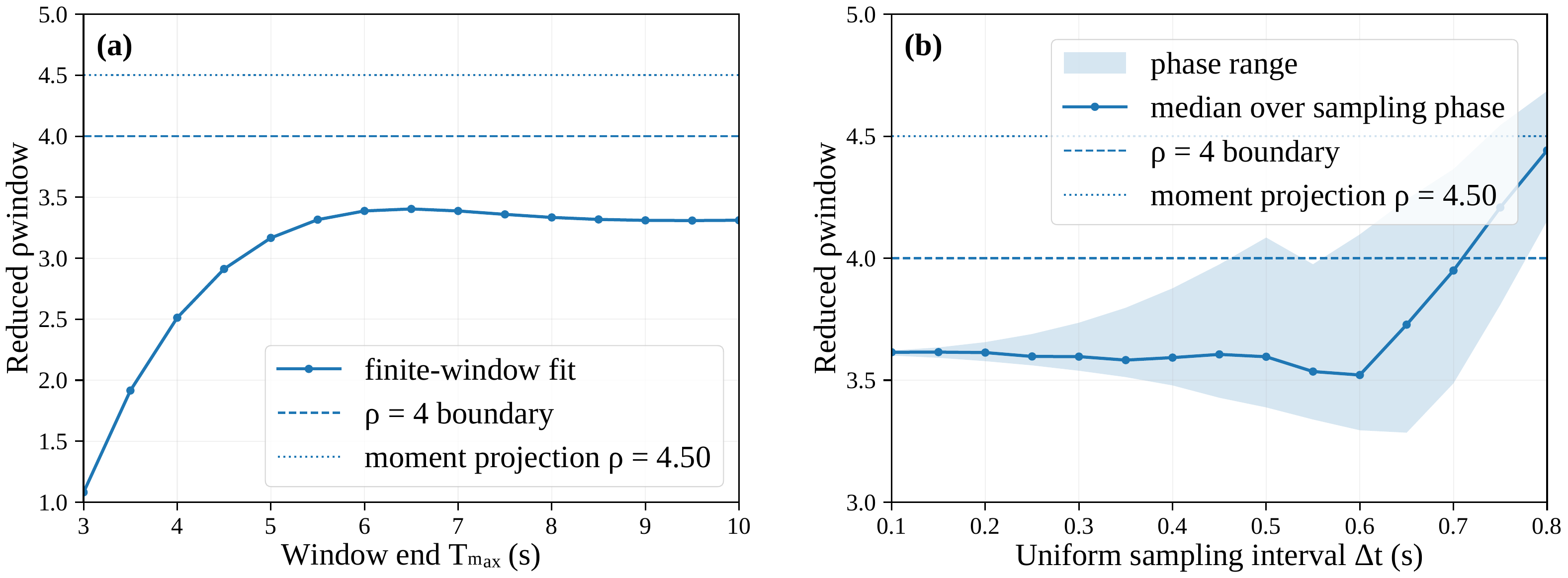}
\caption{Reduction-protocol dependence for one fixed all-real bacterial cascade--parallel realization. (a) The noiseless higher-order response is refitted with the same two-state model after truncating the original graph-digitized sampling schedule at $T_{\max}$. The finite-window estimate stabilizes near $\rho_{\rm window}\simeq3.31<4$, whereas low-frequency moment matching of the same underlying transfer function gives $\rho_{\rm moment}=4.50>4$. (b) At fixed $10$ s duration, the uniform sampling interval $\Delta t$ is varied; the solid curve is the median over sampling phases relative to stimulus onset and the shaded region spans the phase-dependent range. When $\Delta t$ becomes comparable to or exceeds the fast response time, coarse sampling can move the reduced value across $\rho=4$. The figure therefore tests the robustness of Eq.~(\ref{eq:projection}) and shows that the qualitative pole class assigned by a low-order fit depends on how the higher-order system is observed and weighted.}
\label{fig:projectionprotocol}
\end{figure*}

Figure~\ref{fig:projectionprotocol} tests whether the finite-window value is a single cherry-picked number. Truncating the actual bacterial sampling schedule shows that $\rho_{\rm window}$ settles near $3.3$ once the peak and several seconds of relaxation are included; extending the window to the full $10$ s does not drive it toward the moment value because points near the recovered baseline contribute almost no unweighted residual. At fixed $10$ s duration, varying a uniform sampling interval and its phase relative to the stimulus changes the reduced value substantially and can move it across four when the sampling interval becomes comparable to or exceeds the fast response time. Physically, the coarse grid has then lost the information needed to resolve the fastest relaxation mode. Thus the side of the reduced $\rho=4$ boundary is not invariant under observation protocol even for one fixed, noiseless, all-real underlying system.

For comparison, projecting the balanced AWC cascade--parallel trace at its published sampling times yields a real-pole reduced fit close to the direct AWC two-state result (SM). This does not alter the bacterial counterexample: a complex pair in a reduced model is not a unique signature of microscopic phase-space rotation.

\section{Stochastic parallel subtraction}

A deterministic transfer function does not specify how noise enters its internal states. The parallel realization makes this especially transparent. Consider two normalized branch filters driven partly by a common fluctuation $\xi_c$ and partly by independent branch noises $\xi_f$ and $\xi_s$,
\begin{align}
 u_f&=G_f(\xi_c+\xi_f),\qquad
 u_s=G_s(\xi_c+\xi_s),\nonumber\\
 y&=K(u_f-u_s). \label{eq:noisebranches}
\end{align}
For mutually independent white sources of intensities $D_c,D_f,D_s$, the output spectrum is
\begin{align}
 S_y(\omega)=&\;2K^2D_c
 \frac{\omega^2(T_s-T_f)^2}
 {(1+\omega^2T_f^2)(1+\omega^2T_s^2)} \nonumber\\
 &+\frac{2K^2D_f}{1+\omega^2T_f^2}
 +\frac{2K^2D_s}{1+\omega^2T_s^2}+B . \label{eq:parallelPSD}
\end{align}
A common upstream stage multiplies the first term by $|G_a(i\omega)|^2$. Equation~(\ref{eq:parallelPSD}) gives a simple physical decomposition: common-mode noise is rejected at zero frequency by subtraction and therefore produces a finite-frequency band-pass component, while noise generated independently within either branch produces ordinary OU-like low-pass contributions. The same deterministic subtraction can consequently coexist with very different spontaneous spectra, consistent with the general separation between adaptive response and noise filtering in signaling networks \cite{SartoriTu2011}.

Equation~(\ref{eq:parallelPSD}) is a theoretical statement about how noise routing would appear in a parallel realization; it is \emph{not} used to explain the local hills in the present FRET spectra. The released single-cell data of Keegstra \textit{et al.} \cite{Keegstra2017} contain substantial finite-record structure, and three representative spectra are replotted descriptively in the SM with square, triangle, and circle markers. Their smooth curves are guides to the eye chosen to display the broad hill-and-decay shapes, not fits of Eq.~(\ref{eq:parallelPSD}) and not a replacement for the published OU analysis. We therefore make no claim that the visible spectral hills are generated by parallel subtraction. Testing that hypothesis would require longer records and a dedicated stochastic system-identification analysis.

The public data contain 75 finite OU times with mean $12.633$ s and median $11.937$ s. We treat these values as a validation of the published stochastic scale, not as a new measurement. They also belong to a different observable from the tethered-cell motor response in Fig.~\ref{fig:biofits}; no direct fluctuation--response relation is inferred between them. Direct response--fluctuation comparisons in bacterial chemosensory signaling require a matched observable and have been treated separately in single-cell FRET work \cite{Colin2017}.

\subsection{Response and a scalar spectrum do not identify irreversibility}

The original two-state coordinates provide a complementary identifiability result. On the $\kappa=0$ slice, let
\begin{equation}
 A=\begin{pmatrix}-a&-a\\ b&0\end{pmatrix},\qquad
 a=\tau_x^{-1},\quad b=\tau_m^{-1}, \label{eq:Adr}
\end{equation}
and let
\begin{equation}
 D=\begin{pmatrix}D_x&D_c\\D_c&D_m\end{pmatrix} \label{eq:Dmat}
\end{equation}
be the diffusion matrix. The Fourier-domain output is
\begin{equation}
 x(\omega)=\frac{i\omega\eta_x-a\eta_m}{ab-\omega^2+i a\omega}. \label{eq:xFourierMain}
\end{equation}
The two noise transfer amplitudes in the numerator are in exact quadrature. Consequently the real cross-spectral term proportional to $D_c$ cancels identically, and
\begin{equation}
 S_x(\omega)=
 \frac{2\left(D_x\omega^2+a^2D_m\right)}
 {(ab-\omega^2)^2+a^2\omega^2}. \label{eq:Sxmain}
\end{equation}
A recent preprint by Browning \textit{et al.} gives a structural-identifiability analysis of partially observed linear stochastic differential equations, including the two-state OU case \cite{Browning2025}. The same structure is especially transparent in the scalar spectrum. For a general real two-state OU process, with nontrivial coupling, cross diffusion drops out precisely when the hidden state has no self-relaxation. Writing
\begin{equation*}
 A_g=\begin{pmatrix}\alpha&\beta\\ \gamma&\delta\end{pmatrix},
\end{equation*}
and observing only the first state, the cross-diffusion contribution is
\begin{equation*}
 S_x^{(12)}(\omega)=-\frac{4\beta\delta D_{12}}{|\Delta(\omega)|^2},\qquad
 \Delta=(i\omega-\alpha)(i\omega-\delta)-\beta\gamma.
\end{equation*}
Thus, for nontrivial hidden-to-observed coupling ($\beta\neq0$), the scalar spectrum is independent of $D_{12}$ if and only if $\delta=0$. This spectral statement is consistent with the structural-identifiability result of Browning \textit{et al.} \cite{Browning2025}. In the adaptive model it acquires a direct physical interpretation, because the same condition is the integral-memory limit that produces exact adaptation. The adaptive drift in Eq.~(\ref{eq:Adr}) has $\delta=0$, whereas the leaky-memory extension has $\delta=-\kappa b$, so any $\kappa\neq0$ restores sensitivity to cross diffusion. The same parameter that moves the deterministic zero away from the origin therefore also removes the stochastic blind direction. This spectral characterization is specific to the two-state, one-coordinate observation geometry; higher-dimensional systems are governed by the corresponding frequency-dependent transfer-vector geometry.

For the exact-adaptation drift, infinitely many positive-definite diffusion matrices therefore have the same deterministic response and the same scalar $S_x(\omega)$, even though hidden correlations differ. From a statistical-physics viewpoint, this is a marginal-observation problem made exact by integral memory: the scalar spectrum is a two-point statistic of one observed coordinate, whereas irreversibility is encoded by probability currents in the full state space. A hidden noise correlation can therefore change circulation in state space without changing the measured scalar spectrum. Energetic and stochastic-thermodynamic aspects of adaptation have been studied in several complementary frameworks \cite{Lan2012,MehtaSchwab2012,Jia2017,Matsumoto2018,Conti2022}. More general difficulties of inferring irreversibility from partial observations are well known \cite{Martinez2019}; here the spectral form makes the blind direction and its adaptation-specific interpretation explicit.

For the all-even Gaussian time-reversal convention, the stationary path-space irreversibility rate is
\begin{equation}
 \sigma(D_c)=
 \frac{\left[a(D_c+D_m)+bD_x\right]^2}
 {a(D_xD_m-D_c^2)}. \label{eq:sigmaDc}
\end{equation}
Thus $\{H,S_x\}$ does not determine the exact value of $\sigma$: varying the hidden $D_c$ leaves Eq.~(\ref{eq:Sxmain}) unchanged while changing Eq.~(\ref{eq:sigmaDc}). This observational non-identifiability should be kept separate from a second, independent statement. If the \emph{reduced two-state drift} $A$ is taken as given, minimizing over all admissible diffusion matrices yields the drift-only bound
\begin{equation}
 \sigma\ge\frac{1}{\tau_x}\left(\frac{4}{\rho}-1\right),\qquad \rho<4, \label{eq:sigmabound}
\end{equation}
where $\rho$ is the parameter of that reduced two-state drift. Hence a stable focus in this reduced all-even OU model cannot satisfy detailed balance. The result does \emph{not} imply a positive microscopic dissipation bound for the biological system when the same data are better represented by a higher-order all-real realization: the preceding reduction analysis shows precisely why the reduced $\rho<4$ need not be a microscopic invariant.

The all-even convention is natural here only insofar as $x$ and $m$ are interpreted as coarse-grained activity-, occupancy-, or concentration-like biochemical state variables, which are even under time reversal rather than current- or momentum-like variables. If a state were assigned odd parity, the detailed-balance condition and irreversibility functional would change and Eq.~(\ref{eq:sigmabound}) need not hold. The bound is therefore a property of the specified reduced Gaussian model, not a direct measurement of biochemical heat production. Full derivations are given in the SM.

\section{Discussion}

The biological fits and the mathematical results play different roles. Parallel subtraction is not claimed as a newly discovered circuit: positive-minus-negative temporal weighting is established in bacterial chemotaxis, and Kato \textit{et al.} used the same common-stage/parallel-opponent topology for AWC \cite{Segall1986,Kato2014}. Its value here is as a concrete realization that exposes which quantities survive a change of internal representation. Figure~\ref{fig:step} becomes a time-domain picture of transient mismatch between opposed temporal branches, and Fig.~\ref{fig:frequency} is the corresponding frequency-domain statement: common slow components cancel, sufficiently fast components are kinetically attenuated, and intermediate modulation survives. Figure~\ref{fig:biofits} shows that adding the common stage corrects the sharper experimental peaks missed by the minimal model, but that fit improvement is supporting evidence rather than the principal novelty.

The central deterministic result is reduction non-invariance. Figure~\ref{fig:polezero} shows the pole--zero geometry of a specified reduced transfer function, whereas Fig.~\ref{fig:projectionprotocol} shows that the transfer function obtained \emph{after reduction} need not have a protocol-independent pole class. The same fixed all-real cascade--parallel system gives $\rho_{\rm moment}=4.50$ under a low-frequency moment projection and $\rho_{\rm window}=3.31$ under the bacterial finite-window least-squares projection. These are both projections onto the same two-state family; they differ only in which information is weighted. The significant result is therefore not that the numerical optima differ, but that the qualitative classification---real versus complex reduced poles---is reversed. The systematic window and sampling tests in Fig.~\ref{fig:projectionprotocol} show that the finite-window value is not tied to one isolated sampling set and that coarse sampling can itself move the estimate across four.

A separate low-frequency proposition in the SM must not be conflated with this bacterial example. For a positive real-pole cascade, moment matching yields $\rho_{\rm eff}=2N_{\rm eff}/(N_{\rm eff}-1)$, so $N_{\rm eff}>2$ is one analytic route to a complex second-order moment reduction. The fitted bacterial cascade--parallel realization instead has $N_{\rm eff}=1.80$ and $\rho_{\rm moment}=4.50>4$; its $\rho_{\rm window}<4$ arises from finite-window projection, not from the participation-number mechanism. The two mechanisms independently demonstrate that an effective complex pair is not a unique microscopic signature.

The apparent bacterial value $\rho=3.15$ from the graph-digitized two-state fit is correspondingly fragile as a biological classifier. Its predicted undershoot is only $5.6\times10^{-4}$ in motor-bias units, whereas the difference between the fitted baseline and the two prestimulus graph points is about $0.036$, more than sixty times larger. Modest baseline and onset choices move the direct two-state fit from about $\rho=1.67$ to $4.33$ (SM). These sensitivities are not reasons to discard the reduced model; they define what it can and cannot support as a microscopic interpretation.

The stochastic result has the same logical structure at a different level. The general two-state spectral structure is consistent with the recent structural-identifiability analysis of Browning \textit{et al.} \cite{Browning2025}; the present calculation makes its physical meaning in adaptation explicit. In a coupled real two-state OU process observed through one state, cross diffusion is absent from the scalar spectrum exactly when the hidden state has no self-relaxation. The present exact-adaptation model is such an integral-memory system, whereas the leaky-memory parameter $\kappa$ simultaneously moves the deterministic zero away from the origin and restores spectral sensitivity to $D_c$. For $\kappa=0$, Eq.~(\ref{eq:Sxmain}) is therefore blind to $D_c$ while Eq.~(\ref{eq:sigmaDc}) depends strongly on it, so scalar response and scalar fluctuations do not identify path-space irreversibility. The drift-only lower bound in Eq.~(\ref{eq:sigmabound}) is complementary and remains specific to the stipulated reduced all-even adaptive drift; the general spectral condition does not by itself generalize that bound. Because the deterministic part of the paper shows that a reduced focus can itself be an artifact of model reduction, the bound must remain attached to the reduced OU model rather than transferred to the underlying biology.

The supplemental figures retain direct systems meanings. Figure~S1 follows the reduced poles to their coalescence at four; Fig.~S2 shows positive and negative temporal lobes whose equal areas encode dc cancellation; Fig.~S3 shows how leakage separates zero placement from the exceptional-point boundary. Figures~S4--S6 describe the stochastic observation layer. The representative FRET spectra contain visible local hills, but they are shown descriptively and are not used to claim a subtraction-generated stochastic resonance. The published OU times remain the quantitative stochastic summary used here.

\section{Conclusion}

We investigated what can and cannot be inferred about sensory adaptation dynamics from low-dimensional response and fluctuation measurements. Using transfer-function and pole--zero analyses, explicit higher-order realizations, low-frequency moment matching, finite-window system identification, and stochastic Ornstein--Uhlenbeck theory, we examined the effects of internal realization, model reduction, and partial observation on the interpretation of adaptive dynamics. Published \textit{E.~coli} and \textit{C.~elegans} responses were used as biological examples.

We found that different reduction procedures can assign qualitatively different second-order dynamics to a fixed higher-order adaptive system containing only real relaxation modes. Moment matching gives $\rho_{\rm moment}=4.50$. On the other hand, finite-window fitting gives $\rho_{\rm window}=3.31$. Sufficiently coarse sampling can also move the inferred model across the real-complex pole boundary. In the stochastic problem, the general two-state spectrum makes transparent that, with nontrivial coupling and one-state observation, cross diffusion is spectrally invisible precisely when the hidden state has no self-relaxation. In the present adaptive family, this identifiability structure is realized by the integral-memory limit of exact adaptation, while leaky memory restores spectral sensitivity. For the exact-adaptation reduced model, path-space irreversibility nevertheless depends on the hidden cross-diffusion coefficient. Therefore, neither the deterministic response nor the scalar fluctuation spectrum can identify the irreversibility rate in that case. However, specifying an all-even reduced two-state drift with $\rho<4$ imposes the drift-only lower bound $\sigma\ge\tau_x^{-1}(4/\rho-1)$.

These results show that low-order models can reproduce adaptive responses while still misclassifying the inferred relaxation structure or leaving hidden thermodynamic information unidentifiable. They therefore define a practical boundary between what low-dimensional response data can establish and what requires additional state-resolved information.

\end{document}